# Berry's phase in the Josephson phase qubit

Anthony Tyler and Roberto C. Ramos

*Abstract*— Berry's phase often appears in quantum two-level systems with a degeneracy. An example of such a system is a spin-1/2 particle in a magnetic field. As the magnetic field is slowly evolved through a closed path, the particle has been shown to acquire an additional phase called Berry's phase, in addition to the usual dynamical phase. This phase has been found in two-level quantum systems intrinsic to many superconducting qubits and has particularly been calculated for the charge, flux and Josephson phase qubit. Here, we present an alternative derivation of the Berry's phase in a current-biased Josephson junction qubit. We also calculate the complete Berry's phase from the expression $\gamma = i \int \langle\psi|\partial_R\psi\rangle \cdot dR$ evaluated over a closed loop in a frequency parameter space and compare this with the geometric phase. From this comparison, we examine the possibility of using a single phase qubit for topological quantum computing.

## I. INTRODUCTION

Superconducting solid-state qubits represent a particularly promising approach to quantum computing because of their potential for scalability [1]. There are primarily three classes of superconducting qubits: charge-based qubit, flux-based qubit, and phase-based qubit. A major drawback of these qubits is the difficulty in decoupling them from the environment [1]-[2]. To overcome this obstacle, features of these quantum systems that are insensitive to noise, such as their topological phases, are actively being studied. An example of such a topological phase is Berry's phase, which is acquired through adiabatic variations of the Hamiltonian through a closed path [3]. A common example of Berry's phase is a spin-1/2 particle in a magnetic field. As the magnetic field is adiabatically varied, so is the Hamiltonian. This results in the spin state following the magnetic field direction. When the magnetic field returns to its initial position, the spin state has acquired the usual dynamical phase and an extra phase called Berry's phase that depends on the actual path taken (See Fig. 1).

It is easiest to determine the Berry's phase of a solid state qubit if we model it as a two-level system that is analogous to the spin-1/2 particle in a magnetic field. This allows us to define an analogous magnetic field which will be adiabatically varied. Using this approach, the Berry's phase of single charge and flux qubits have been calculated [4]-[5]. More recently, the Berry's phase of a current-biased Josephson junction was explored in [6]. Here, we present an alternative derivation of this result. Additionally, we explore the possibility of using this phase to perform topological quantum computing, which is a form of quantum computation that operates upon quantum systems through meta-particles known as anyons. Anyons arise in 2 + 1D braiding statistics. Of particular interest are non-abelian anyons. In this case, when a braiding operation is performed on two of them, the same operation with anyons switched will result in a different outcome. For the phase qubit to be a candidate for this form of computation, it must satisfy four conditions: First, its Berry's phase $\gamma = \gamma_g + \alpha$ must have a geometric portion $\gamma_g$, which is proportional to the solid angle enclosed by the closed path, and a non-zero, non-geometric portion $\alpha$. Fulfillment of this condition means that anyons can be found. The three other conditions are that the ground state is nearly degenerate, that adiabatic exchanges of anyons results in a unitary transformation being applied to the ground state, and that this transformation is only applied when anyons are exchanged. If these four conditions are met, then the anyons are said to follow non-abelian braiding statistics. [7]

## II. THEORY

The Hamiltonian of the phase qubit can be written in the form:

$$H = E_C N^2 - E_J(\cos\varphi + (I/I_o)\varphi) \quad (1)$$

where N is the number of Cooper pairs on the junction; $E_C = (2e)^2/2C_J$, the charging energy; $E_J = \hbar I_o/2e$, the Josephson coupling energy; $\varphi$ is the phase across the junction; $e$ is the electronic charge; I is the bias current; $I_o$ is the junction's critical current; and $C_J$ is the junction capacitance. This can be converted to a spin representation:

$$H = -(\hbar\omega_p/2)\sigma_z \quad (2)$$

where $\omega = \{\sqrt{[2E_JE_C]}/\hbar\}(1-(I/I_o)^2)^{1/4}$ is the frequency of the energy spacing of the two lowest energy levels and $\sigma_z$ is a Pauli spin matrix. To account for time-dependent currents in general, a change of basis gives us:

$$H = -(\hbar\omega_p/2)\sigma_z - (E_J l/\sqrt{2})(I(t) - I_B)/I_o)\sigma_x \quad (3)$$

where $I_B$ is the time-independent bias current; I(t) is the time-dependent current; $l = (2E_C/E_J)^{1/4}(1-I_B/I_o)^{1/8}$ is a dimensionless constant of the system resulting from the change of basis. For this form of Hamiltonian to hold, it is necessary that $(I(t) - I_B)/I_o \ll 1$ [2].

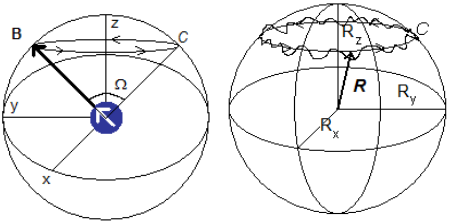

Fig. 1 Left: As the magnetic field is slowly varied around a closed loop, the spin state closely follows. When the magnetic field returns to its original position, the spin state would have acquired both a dynamical phase and an extra Berry's phase that is proportional to the solid angle subtended by the loop. Right: Topological phases are relatively insensitive to noise from the environment.

## III. DERIVATION OF BERRY'S GEOMETRIC PHASE AND THE COMPLETE BERRY'S PHASE

Our alternative derivation of Berry's phase begins with the Hamiltonian in Equation (3). In order to meet the condition $(I(t) - I_B)/I_o \ll 1$, we apply a microwave signal $I(t) = I\cos(\omega t + \delta) + I + I_B$, where I is the amplitude of an applied microwave signal; $\omega$ is the frequency; and $\delta$ is the phase. This results in:

$$H = -(\hbar\omega_p/2)\sigma_z - (E_J lI/(\sqrt{2}I_o))(\cos(\omega t + \delta) + 1)\sigma_x \quad (4)$$

from which we can define a parameter space vector:

$$\mathbf{R} = (E_J lI/(\sqrt{2}I_o))(\cos(\omega t + \delta) + 1)\,\mathbf{i} + \hbar\omega_p\,\mathbf{k} \quad (5)$$

that clearly oscillates parallel to the x-axis. To remove the time dependence and allow for adiabatic evolution of the parameters of the applied microwave signal, we apply the rotating frame approximation. By rotating our reference frame by the angle

$$\theta = \arctan\left(\frac{\hbar\omega_p}{\sqrt{\left(\frac{E_J lI \sqrt{2}}{I_0}\right)^2 + (\hbar\omega_p)^2}}\right) \quad (6)$$

which transforms the Hamiltonian to $H_r$. Applying the rotating frame approximation

$$H_{eff} = \exp\left(\frac{i\omega t}{2}\sigma_z\right) H' \exp\left(-\frac{i\omega t}{2}\sigma_z\right) + \frac{\hbar\omega}{2}\sigma_z \quad (7)$$

we find that

$$H_{eff} = -\frac{E_J lI}{2\sqrt{2}I_0}\cos\phi\,\sigma_x - \frac{E_J lI}{2\sqrt{2}I_0}\sin\phi\,\sigma_y + \frac{\hbar\omega - \hbar\omega_p}{2}\sigma_z \quad (8)$$

We can now define a three-dimensional parameter space vector

$$\vec{R} = \frac{E_J lI}{\sqrt{2}I_0}\cos\phi\,\hat{i} + \frac{E_J lI}{\sqrt{2}I_0}\sin\phi\,\hat{j} + (\hbar\omega_p - \hbar\omega)\,\hat{k} \quad (9)$$

from which we see that by adiabatically varying the phase $\delta$ of the applied microwaves from 0 to $2\pi$, we can trace a closed path in parameter space defined by:

$$z = \hbar\omega_p - \hbar\omega \qquad r = \frac{E_J lI}{\sqrt{2}I_0} \quad (10)$$

From Ref. [3], the Berry's phase $\gamma$ of the system should be proportional to the solid angle $\Omega$ subtended by the closed path. Since $\mathbf{C} = \partial H_{eff}/\partial \mathbf{R} = -\boldsymbol{\sigma}/2$, we find:

$$\gamma = C\Gamma = -\Omega/2 = \pi[1 - z/\sqrt{(z^2 + r^2)}] \quad (11)$$

where $z = \hbar\omega_p - \hbar\omega$ and $r = E_J lI/\sqrt{2}I_o$. Comparing this with the results in Ref. [6], they are found to be identical.

To calculate the *complete* Berry's phase, we first calculate the states of $H_{eff}$. The energies and corresponding states of this $H_{eff}$ are found to be:

$$E_0 = +\frac{1}{2}\sqrt{z^2 + r^2}$$
$$E_1 = -\frac{1}{2}\sqrt{z^2 + r^2}. \quad (12)$$

with the corresponding states:

$$|0\rangle = \begin{pmatrix} \frac{(z-E_0)(\cos\phi - i\sin\phi)}{\sqrt{2r^2 + 2z^2 - 2zE_0}} \\ \frac{r}{\sqrt{2r^2 + 2z^2 - 2zE_0}} \end{pmatrix}$$

$$|1\rangle = \begin{pmatrix} \frac{(z-E_1)(\cos\phi - i\sin\phi)}{\sqrt{2r^2 + 2z^2 - 2zE_1}} \\ \frac{r}{\sqrt{2r^2 + 2z^2 - 2zE_1}} \end{pmatrix}. \quad (13)$$

Using these results and the prescription for the complete Berry's phase:

$$\gamma = i\oint \langle\psi|\nabla_{\vec{R}}\psi\rangle \cdot d\vec{R} \quad (14)$$

we find that the complete Berry's phase is identical to the result in Equation (11). Because there is no non-geometric part of the complete Berry's phase, this implies that single phase qubits do not exhibit any anyonic behavior and by themselves, cannot be used for topological quantum computing.

## IV. CONCLUSION

We have calculated the complete Berry's phase $\gamma = \gamma_g + \alpha$ for the Josephson phase qubit where $\gamma_g$ is the geometric portion derived in [6] and alternatively derived here and $\alpha$ is the non-geometric portion. Since $\alpha$ is zero, we have shown that single phase qubits do not exhibit any anyonic behavior and cannot be used for topological quantum computing. We are currently extending this search to multiple phase qubits as well as qubits coupled to harmonic oscillators.

REFERENCES

[1] M.H. Devoret and J. M. Martinis, *Quantum Information Processing* **3**, 163-203 (2004)
[2] M.R. Geller and E.J. Pritchett, "Quantum Computing with Superconductors I: Architectures", e-print quant-ph/0603224v1 (2006)
[3] M.V. Berry, "Quantal phase factors accompanying adiabatic changes'", *Proc. R. Soc. London*, Ser. A. **292**, 45 (1984)
[4] Z.H. Peng, M.J. Zhang and D.N. Zheng, "Detection of geometric phases in flux qubits with coherent pulses", *Phys. Rev. B* **73**, 020502 (2006)
[5] P.J. Leek, et al., "Observation of Berry's Phase in a Solid-State Qubit", *Science* **318**, 1889-1892 (2007)
[6] Z.H. Peng, F Chu, Z.D. Wang and D.N. Zheng, "Implementation of adiabatic geometric gates with superconducting phase qubits" *J. Phys.: Cond. Matt* **21**, 045701 (2009)
[7] S. Das Sarma, et al., e-print cond-matt/0707.1889v2 (2008)